\documentclass[prb,twocolumn,showpacs,preprintnumbers,amsmath,floatfix]{revtex4}
\usepackage{graphicx}

\begin{document}

\title{Microscopic origin of pressure-induced  phase transitions in iron-pnictide AFe$_{2}$As$_{2}$ superconductors:
 an \textit{ab initio} molecular-dynamics study }
\author{Yu-Zhong Zhang}
\affiliation{Institut f{\"u}r Theoretische
Physik, Goethe-Universit{\"a}t Frankfurt, Max-von-Laue-Stra{\ss}e 1,
60438 Frankfurt am Main, Germany}
\author{Hem C. Kandpal}
\affiliation{Institut f{\"u}r Theoretische
Physik, Goethe-Universit{\"a}t Frankfurt, Max-von-Laue-Stra{\ss}e 1,
60438 Frankfurt am Main, Germany}
\author{Ingo Opahle}
\affiliation{Institut f{\"u}r Theoretische
Physik, Goethe-Universit{\"a}t Frankfurt, Max-von-Laue-Stra{\ss}e 1,
60438 Frankfurt am Main, Germany}
\author{Harald O. Jeschke}
\affiliation{Institut f{\"u}r Theoretische
Physik, Goethe-Universit{\"a}t Frankfurt, Max-von-Laue-Stra{\ss}e 1,
60438 Frankfurt am Main, Germany}
\author{Roser Valent{\'\i}}
\affiliation{Institut f{\"u}r Theoretische
Physik, Goethe-Universit{\"a}t Frankfurt, Max-von-Laue-Stra{\ss}e 1,
60438 Frankfurt am Main, Germany}
\date{\today}

\begin{abstract}
  Using {\it ab initio} molecular dynamics we investigate the
  electronic and lattice structure of $A$Fe$_2$As$_2$ ($A$=Ca, Sr, Ba)
  under pressure. We find that the structural phase transition
  (orthorhombic to tetragonal symmetry) is always accompanied by a
  magnetic phase transition in all the compounds, while the nature of
  the transitions is different for the three systems. Our calculations
  explain the origin of the existence of a  collapsed tetragonal phase in CaFe$_2$As$_2$
  and its absence in BaFe$_2$As$_2$. We argue that changes of the Fermi surface
   nesting features dominate the phase transitions under pressure rather than
   spin frustration or a Kondo scenario. The consequences for superconductivity
  are discussed.
\end{abstract}


\maketitle

The discovery of iron pnictide superconductors~\cite{Kamihara} with
critical temperatures $T_c$ up to 57.4~K~\cite{Cheng} upon doping
has strongly revived the interest in high-T$_c$ superconductivity.
The undoped Fe-based parent compound undergoes at low temperatures a
structural transition from tetragonal to orthorhombic symmetry
accompanied by a magnetic phase transition to a stripe-type
spin-density wave state~\cite{Cruz,Huang,Rotter,Goldman,Zhao}. While
the nature of these two transitions is different between LaFeAsO
(1111 compound) and $A$Fe$_2$As$_2$ (122 compound) with $A$=(Ba, Sr,
Ca), superconductivity appears in both material classes only when
the lattice distortion and magnetic ordering are suppressed,
indicating a strong competition between the structural distortion,
magnetic ordering and superconductivity in iron pnictides.

Recently, superconductivity in the parent compounds 1111 and 122 was
reported under application of pressure
~\cite{Okada,Igawa,Park,Torikachvili,Kreyssig,Goldman2,Pratt,Yu,Lee,Kumar,Kotegawa,Kotegawa2,Torikachvili2,Fukazawa,Alireza,Kimber,Colombier,Mani}.
In LaFeAsO~\cite{Okada}, resistivity measurements show
superconductivity at $\approx 12$~GPa with $T_c=21$~K. In
BaFe$_2$As$_2$ superconductivity is found to appear gradually with
increasing pressure while in SrFe$_2$As$_2$ the onset of
superconductivity occurs abruptly~\cite{Alireza}. In
CaFe$_2$As$_2$~\cite{Kreyssig,Goldman2,Pratt}, detailed neutron and
X-ray diffraction analysis shows that the system  undergoes
a first order  phase transition from a magnetic orthorhombic to
a nonmagnetic 'collapsed' tetragonal phase under pressure. The possible
appearance of superconductivity in this 'collapsed' tetragonal phase is presently
under debate~\cite{Yu,Lee}.
 While various experiments give
different values of critical
pressures~\cite{Kumar,Kotegawa,Kotegawa2,Torikachvili2,Fukazawa,Alireza,Kimber,Colombier,Mani}
due to the fact that the phase transition is sensitive to possible
nonhydrostatic pressure effects,  Sn content in some samples, or the
use of single crystals or polycrystalline material for structure determination, it is claimed that BaFe$_2$As$_2$ and
SrFe$_2$As$_2$ do not manifest a 'collapsed' tetragonal phase at
elevated pressure~\cite{Goldman2,Kumar,Kotegawa,Kimber,Colombier}.
Therefore, the fact that structurally similar compounds exhibit
phase transitions of different nature urgently calls for a
theoretical understanding. Moreover, it is still under intensive debate
 which is the driving mechanism of the collinear stripe-type
antiferromagnetic ordering;  whether the Fermi surface nesting
or the competition of exchange antiferromagnetic interactions
between the nearest neighbor and next-nearest neighbor
irons~\cite{Singh,Fang,Dong,Mazin2,Yildirim2,Si,Ma,Han,Ma2,Ma3,Han2,Zhao1}.

Theoretical work on optimization of cell parameters and atomic
positions under pressure within the framework of density functional
theory (DFT) has been done on $A$Fe$_2$As$_2$ ($A$=Ca, Ba,
Sr)~\cite{Yildirim,Xie,Kasinathan09} and
LaOFeAs~\cite{Yildirim,Opahle}. However, by considering
Vanderbilt-type ultrasoft pseudopotentials, Yildirim~\cite{Yildirim}
obtained a smooth structural transition for CaFe$_2$As$_2$ under
pressure, observing neither a sudden sizable increase of the cell
parameters $a$ and $b$ nor a strong decrease of the cell parameter
$c$ which is inconsistent with experimental results~\cite{Kreyssig}.
Xie {\it et al.}~\cite{Xie} optimized within the full potential
linearized augmented plane wave method (FPLAPW) the orthorhombic
lattice structure for BaFe$_2$As$_2$ under pressure by relaxing the
internal parameter $z_\text{As}$ and the $c/a$ ratio while keeping
the $b/a$ ratio fixed. This procedure doesn't allow for the
detection of the structural and magnetic phase transitions. Opahle
{\it et al.}~\cite{Opahle} investigated LaOFeAs under pressure and
found that the system is close to a magnetic instability. Obviously,
a complete and unambiguous theoretical description of the
pressure-induced phase transitions in $A$Fe$_2$As$_2$ ($A$=Ca, Sr,
Ba) is still missing.

In this paper, we employ the Car-Parrinello~\cite{CarParrinello}
projector-augmented wave~\cite{Bloechl} molecular dynamics method at
constant pressure~\cite{Rahman} in order to investigate the
pressure-induced phase transitions for $A$Fe$_2$As$_2$ ($A$=Ca, Sr,
Ba). Since in such transitions the interplay among electronic,
magnetic and lattice dynamics is essential, a combined {\it ab initio}
DFT with molecular dynamics approach as used in the present work is
very suitable. In such a procedure, at each pressure value a full,
unbiased relaxation of all lattice and electronic degrees of freedom
is performed in time steps of 0.12~fs at zero temperature.  We used
$4\times 4 \times 4$ $k$-points in doubled
($\sqrt{2}\times\sqrt{2}\times 1$) unit cells.
We used high energy cutoffs of 612~eV and 2448~eV for the wave
functions and charge density expansion respectively. The total energy
was converged to less than 0.01~meV/atom and the cell parameters to
less than 0.0005~\AA. The 3s3p3d (4s4p4d/5s5p5d) states in Ca (Sr/Ba)
and the 3d4s4p states in Fe and As are treated as valence states. We
checked our calculations with the FPLAPW method as implemented in the
WIEN2k code~\cite{Blaha}. Very good agreement is found between these
two methods. The Perdew-Burke-Ernzerhof generalized gradient
approximation (GGA) to DFT has been used.

\begin{figure}[tb]
\includegraphics[width=0.45\textwidth]{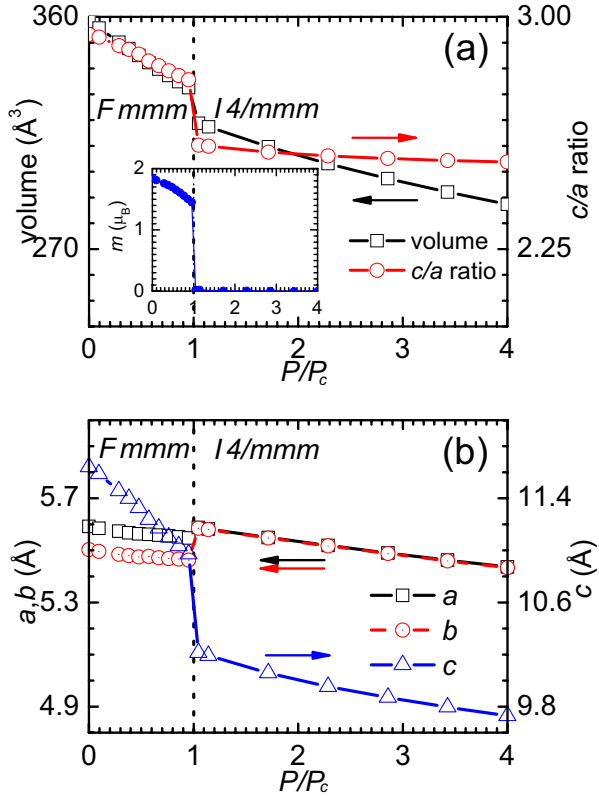}
\caption{(Color online) Calculated changes of (a) volume, $c/a$
ratio and magnetization (inset), (b) lattice parameters as a
function of applied external pressure normalized to the critical
pressure for CaFe$_2$As$_2$. The phase boundary between $F\,mmm$ and
$I\,4/mmm$ is indicated by the vertical dashed line. }
\label{fig:Ca1}
\end{figure}

\begin{figure}[tb]
\includegraphics[width=0.40\textwidth]{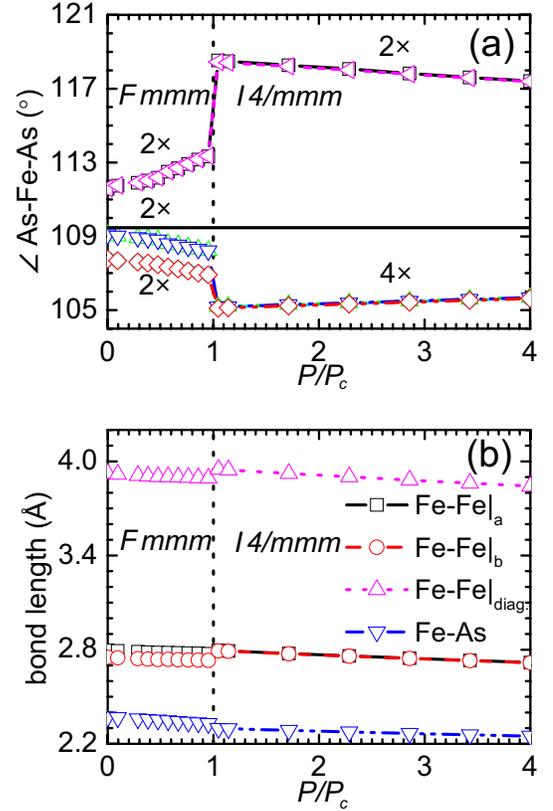}
\caption{(Color online) Calculated changes of (a) As-Fe-As angles,
and (b) Fe-Fe, Fe-As bond length as a function of applied external
pressure normalized to the critical pressure for CaFe$_2$As$_2$. The
phase boundary between $F\,mmm$ and $I\,4/mmm$ is indicated by the
vertical dashed line. } \label{fig:Ca2}
\end{figure}

Our findings can be summarized as follows: We show that
CaFe$_2$As$_2$ undergoes a first order phase transition to a
collapsed tetragonal phase with relative changes of lattice
parameters, bond lengths and angles agreeing very well with the
experimental observations.  In particular our calculations can
account for the measured expansion along the $ab$
plane~\cite{Kreyssig} at the critical pressure, which was not
obtained in previous calculations. We also find an abrupt
disappearance of magnetization at the critical pressure.  For
SrFe$_2$As$_2$ and BaFe$_2$As$_2$, where less is known about the
details of the  lattice structure and  magnetization changes under
pressure, we obtain a simultaneous structural (orthorhombic to
tetragonal) and magnetic phase transition at high pressure. However,
we observe a weak first-order phase transition for SrFe$_2$As$_2$
and a continuous phase transition for BaFe$_2$As$_2$ in contrast to
the strongly first order phase transition in CaFe$_2$As$_2$.  We
explain the microscopic origin of these differences in terms of the
different Fermi surface behavior under pressure. Finally, we argue
that the existence of a magnetic transition from a striped AF state
to a non-magnetic state under pressure is mainly driven by changes
of the Fermi surface nesting rather than spin frustration or a Kondo
scenario~\cite{Vojta}.

\begin{figure}[tb]
\includegraphics[width=0.45\textwidth]{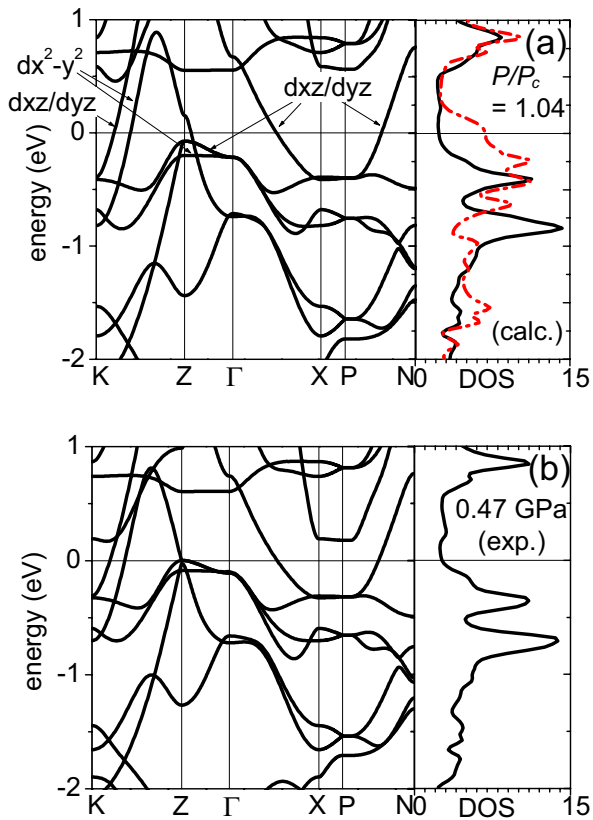}
\caption{(Color online) Comparison of the CaFe$_2$As$_2$ band
structure and DOS in the 'collapsed' tetragonal phase obtained from
(a)
 our optimized lattice structure at $P/P_c = 1.04$ and (b)
the experimental lattice structure~\protect\cite{Kreyssig} at
$P^{\rm exp}/P^{\rm exp}_c \approx 1.6$.  In (a) the total DOS for
the high $T$ tetragonal phase is also shown by the dashed (red)
line.} \label{fig:com} \end{figure}

In Figs.~\ref{fig:Ca1}~(a) and (b) we present the calculated changes
of the volume and lattice parameters of CaFe$_2$As$_2$ as a function
of pressure.  The volume and the lattice parameter $c$ decrease
gradually with increasing pressure and show a discontinuous
shrinkage at the critical pressure, where the system undergoes a
structural phase transition from orthorhombic symmetry to a volume
'collapsed' tetragonal symmetry.  Our results are in very good
agreement with experimental data~\cite{Kreyssig} with a volume
collapse of $\Delta V^{\rm th} \approx 4.1\%$, $\Delta V^{\rm exp}
\approx 4\%$. Moreover, the magnetization sharply goes to zero as
shown in the inset of Fig.~\ref{fig:Ca1}~(a). Surprisingly, however,
the lattice parameters $a$ and $b$ show abrupt expansions at the
phase boundary although otherwise they show a monotonous compression
in both phases. Such expansions, as also observed in
experiments~\cite{Kreyssig}, were not detected in previous DFT
calculations and have a fundamental physical origin. In the
orthorhombic phase below the critical pressure, the system shows
collinear AF ordering within $ab$ plane. At the critical pressure,
the AF ordering is destroyed and the Pauli principle becomes more
effective due to the appearance of more parallel Fe spins giving
rise to a non-negligible lattice expansion along $a$ and
$b$~\cite{ZJV}. As a consequence, the Fe-Fe bond length increases at
the phase transition as observed in Fig.~\ref{fig:Ca2}~(b). In
contrast, the As atoms move towards the Fe plane at the phase
transition and the Fe-As bond length suddenly decreases with a
relative shrinkage of $\Delta d_{Fe-As}^{\rm th} \approx 1.3 \%$ in
very good agreement with $\Delta d_{Fe-As}^{\rm exp} \approx 1.2
\%$. Accordingly, the angles of the tetrahedra are suddenly shifted
away from the ideal tetrahedral value of 109.47$^\circ$ at the phase
transition, while in the orthorhombic and in the high temperature
tetragonal phase they are much closer to the ideal value
(Fig.~\ref{fig:Ca2}~(a)). The strong distortion of the tetrahedron
enhances the crystal field splitting between $d_{x^2-y^2}$,
$d_{xz}$, $d_{yz}$ and $d_{z^2}$, $d_{xy}$ (orbitals given in the
$x\parallel a$, $y\parallel b$, $z\parallel c$ local reference
frame).  This can be seen in the density of states (DOS) shown in
Fig.~\ref{fig:com}~(a) where the two peaks just below (of
$d_{x^2-y^2}$, $d_{xz}$, $d_{yz}$ character) and above (of
$d_{z^2}$, $d_{xy}$ character) the Fermi level in the high
temperature ambient pressure tetragonal phase (dashed (red) line)
are further separated in the high pressure volume 'collapsed'
tetragonal phase (solid (black) line).

\begin{figure}[tb]
\includegraphics[width=0.45\textwidth]{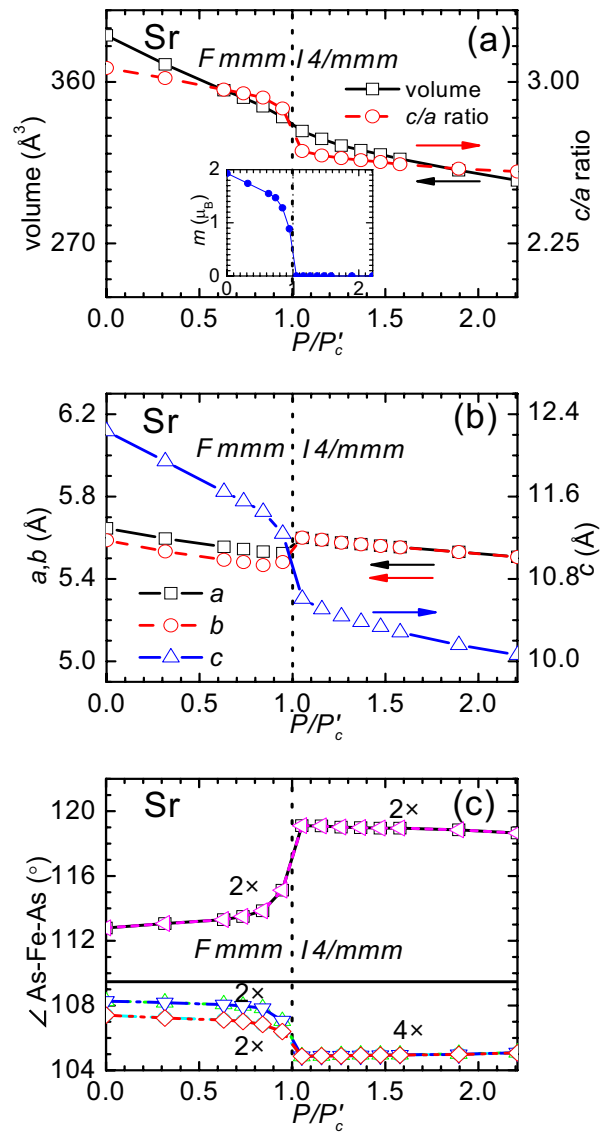}
\caption{(Color online) Calculated changes of (a) volume, $c/a$
ratio and magnetization (inset), (b) lattice parameters and (c)
As-Fe-As angles as a function of applied external pressure
normalized to the critical pressure for SrFe$_2$As$_2$. The phase
boundary between $F\,mmm$ and $I\,4/mmm$ is indicated by the
vertical dashed line.} \label{fig:Sr}
\end{figure}

\begin{figure}[tb]
\includegraphics[width=0.45\textwidth]{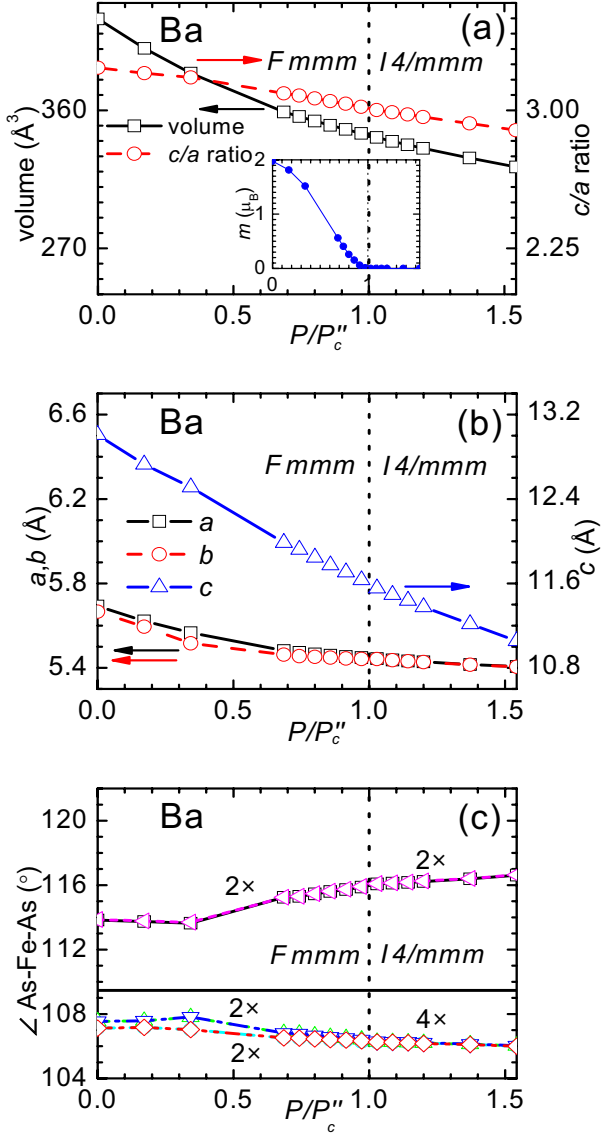}
\caption{(Color online) Calculated changes of (a) volume, $c/a$
ratio and magnetization (inset), (b) lattice parameters and (c)
As-Fe-As angles as a function of applied external pressure
normalized to the critical pressure for BaFe$_2$As$_2$. The phase
boundary between $F\,mmm$ and $I\,4/mmm$ is indicated by the
vertical dashed line. } \label{fig:Ba}
\end{figure}

One aspect of the calculations to be considered is that the
calculated critical pressure $P_c=5.25 \pm 0.25$~GPa is an order of
magnitude larger than the experimental critical pressure $P_c^{\rm
exp} \approx 0.3$~GPa. The overestimation of $P_c$ in our
calculations is a consequence of the well-known overestimation of
volume and magnetic moment by GGA in the Fe pnictides at ambient
pressure~\cite{Opahle,Mazin}. Therefore in order to reach the
experimental structure conditions we have to go to higher simulation
pressures. This procedure is proven to be valid since, as discussed
above, we can reproduce the experimentally reported structural
features of the phase transition in CaFe$_2$As$_2$, as well as the
electronic properties in both orthorhombic phase~\cite{ICM2009} (not
shown) and 'collapsed' tetragonal phase as shown in
Figs.~\ref{fig:com}~(a) and (b) where we present the comparison
between the electronic structure obtained from the
Car-Parrinello-derived lattice structure (at $P/P_c=1.04$) and that
obtained from the experimentally measured one (at 0.47 GPa, $P^{\rm
exp}/P_c^{\rm exp} \approx 1.6$). We observe that the shapes of the
total DOS are almost the same and only slight differences are found
in details of the band structure. Therefore we can conclude that,
although our calculations always overestimate the critical pressure,
the phase transition and the physical properties in both phases can
be quantitatively captured by our analysis.

\begin{figure}[tb] \includegraphics[width=0.45\textwidth]{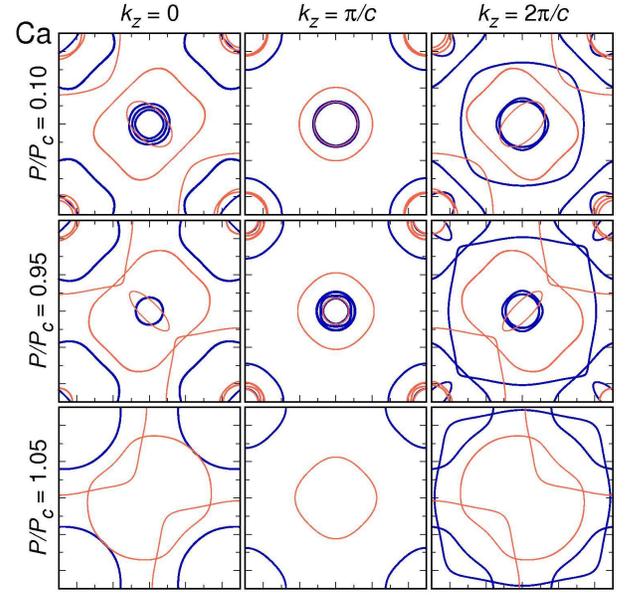}
\caption{(Color online) Fermi surface cuts centered at (0,0,$k_z$)
  with $k_z$ = 0, $\pi/c$ and $2 \pi/c$ and $\left| {k_x} \right|$,
  $\left| {k_y} \right| \leq \pi/a$ at three pressure values for
  CaFe$_2$As$_2$. The Fermi surface sheets are shown in black (blue).
  The grey (orange) sheets result from translating the sheets centered
  at ($\pi/a,\pi/a, k_z$) by ($-\pi/a,-\pi/a,0$) in order to see the
  nesting effects at (0,0,$k_z$). Note, that the low $P$ Fermi surface
  should not be compared to ambient $P$ high $T$ Fermi surface as we
  always work at zero temperature. Here the pressures are normalized
  with respect to the critical pressure $P_c$ for
  CaFe$_2$As$_2$.} \label{fig:BSFS1}
\end{figure}

In Fig.~\ref{fig:Sr} and Fig.~\ref{fig:Ba}, we show the results for
SrFe$_2$As$_2$ and BaFe$_2$As$_2$.  We find that the structure and
the magnetic phase transitions under pressure still occur
simultaneously in these two compounds, but the nature of the phases
is distinctly different from the CaFe$_2$As$_2$ case. SrFe$_2$As$_2$
shows smaller magnetization, volume, lattice and angles jumps at the
critical pressure (see Fig.~\ref{fig:Sr}) compared to CaFe$_2$As$_2$
while we can hardly detect any discontinuity in BaFe$_2$As$_2$ (see
Fig.~\ref{fig:Ba}). The simultaneous and abrupt change of the
lattice structure and the magnetization in SrFe$_2$As$_2$ is again
consistent with recent experimental findings~\cite{Kumar,Kotegawa}
while our results in BaFe$_2$As$_2$ have not yet been completely
confirmed experimentally. Here, the pressure is again normalized to
the critical simulated pressure $P_c'=9.5 \pm 0.5$~GPa for
SrFe$_2$As$_2$ and $P_c''=17.5 \pm 0.5$~GPa for BaFe$_2$As$_2$. The
calculated critical pressure for SrFe$_2$As$_2$ and BaFe$_2$As$_2$
is about 2-3 larger than the experimental one, while for
CaFe$_2$As$_2$ it is around 20 times larger. This can be understood
from the already mentioned overestimation of magnetic interactions
and volumes in GGA calculations. Due to this, one would expect an
(approximately constant) shift between the calculated and measured
$P_c$, thus resulting in the large deviation for CaFe$_2$As$_2$
(where $P_c$ is small) and the better agreement for SrFe$_2$As$_2$
and BaFe$_2$As$_2$.

\begin{figure}[tb] \includegraphics[width=0.45\textwidth]{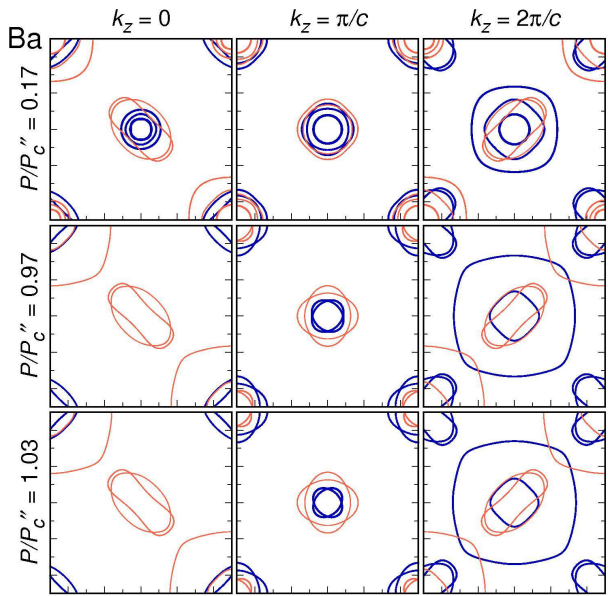}
\caption{(Color online) Fermi surface cuts centered at (0,0,$k_z$)
  with $k_z$ = 0, $\pi/c$ and $2 \pi/c$ and $\left| {k_x} \right|$,
  $\left| {k_y} \right| \leq \pi/a$ at three pressure values for
  BaFe$_2$As$_2$. The color coding is the same as in
  Fig.~\ref{fig:BSFS1}. Here the pressures are normalized with respect
  to the critical pressure $P_c''$ for
  BaFe$_2$As$_2$.} \label{fig:BSFS2}
\end{figure}

\begin{figure}[tb] \includegraphics[width=0.45\textwidth]{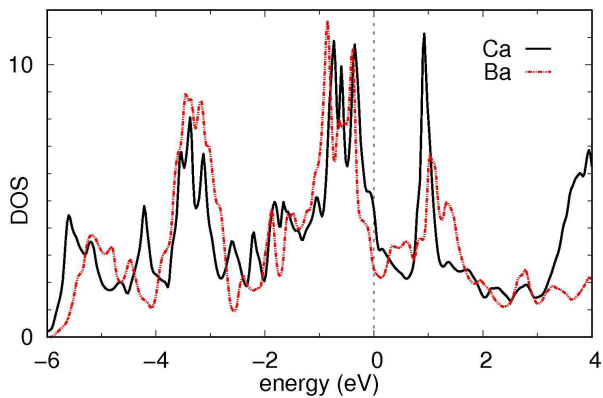}
\caption{(Color online) Total DOS for CaFe$_2$As$_2$ (continuous
line) and BaFe$_2$As$_2$ (dashed line) at the critical pressure
region. } \label{fig:CaBaDOS}
\end{figure}

In order to understand the microscopic origin of the differences in behaviour between SrFe$_2$As$_2$, BaFe$_2$As$_2$ and CaFe$_2$As$_2$,
we performed non-spin-polarized calculations with the optimized
lattice structures and analyzed the band structures and Fermi surfaces.  Below
the critical pressure, we considered an average of lattice
parameters $a$ and $b$ as a single lattice parameter in order to
make the comparison of band structures of different phases
(orthorhombic and tetragonal) possible within the same Brillouin
zone.  It has already been noticed that the tiny differences between
$a$ and $b$ has no appreciable influence on the electronic
structure~\cite{Opahle}.

\begin{figure}[tb]
\includegraphics[width=0.35\textwidth]{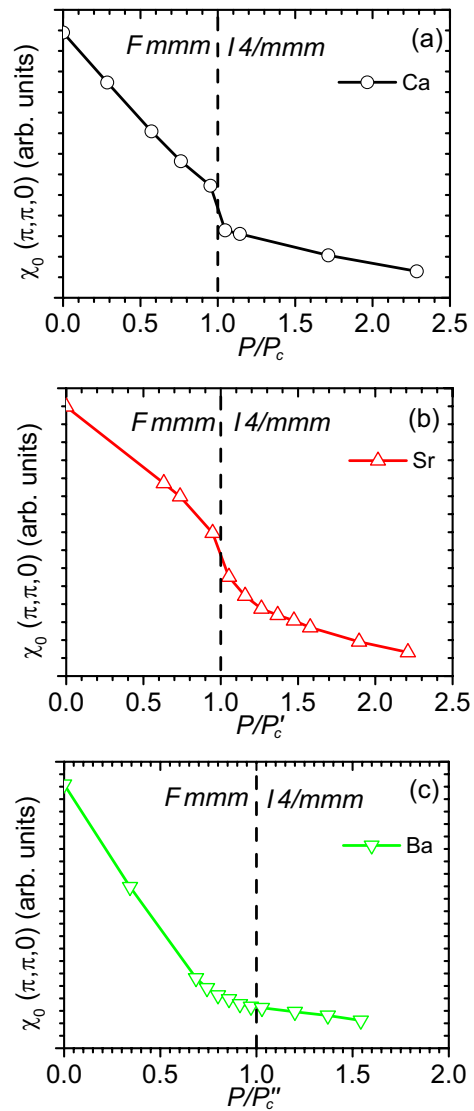}
\caption{(Color online) Calculated noninteracting susceptibility
$\chi_0 (q)$ at $q=(\pi,\pi,0)$ within the constant matrix element
approximation as a function of the normalized pressure with respect
to the critical one (in arbitrary units). This magnitude  measures
the response for the observed stripe-type antiferromagnetic
ordering. The three panels correspond to: (a) CaFe$_2$As$_2$, (b)
SrFe$_2$As$_2$, and (c) BaFe$_2$As$_2$. The phase boundary between
$F\,mmm$ and $I\,4/mmm$ symmetries is indicated by the vertical
dashed line. } \label{fig:SUS}
\end{figure}

In Figs.~\ref{fig:BSFS1} and \ref{fig:BSFS2} we show $k_z$ =0, $\pi/c$,
and $2\pi/c$
 Fermi surface cuts of CaFe$_2$As$_2$
and BaFe$_2$As$_2$ at three different pressures.  As a function of
pressure, the changes of the Fermi surface in both compounds is
remarkably different. Below $P_c$, CaFe$_2$As$_2$ shows with
increasing pressure gradual shrinkages of the  nested inner hole and
electron cylinders around (0,0,$k_z$) and ($\pi/a,\pi/a,k_z$)
respectively leading to a reduction of the nesting features  at each
$k_z$ (compare in Fig.~\ref{fig:BSFS1} the rows at $P/P_c$=0.1 and
$P/P_c$=0.95) and consequently a continuous reduction of the
magnetization in the orthorhombic phase (inset of
Fig.~\ref{fig:Ca1}~(a)). As pressure increases to the critical
region, the residual steep DOS at the Fermi level
(Fig.~\ref{fig:CaBaDOS}) still forces the system to find a way to
alleviate such an instability. However, the nested inner cylinders
become too small to remove it by magnetic ordering.
 A sudden distortion of the
tetrahedron is realized to lower the electronic energy by further
splitting the Fe~$3d$ orbitals (see the dashed (red) DOS line in
Fig.~\ref{fig:com} (a)) at the price of increasing the lattice
energy. Thus, as shown in Fig.~\ref{fig:BSFS1},
$P/P_c$=1.05 row, right above the critical pressure all the
cylinders around (0,0,$k_z$) completely vanish due to the splitting
of Fe~$3d$ orbitals by lattice distortions in contrast to
BaFe$_2$As$_2$ (see below), and accordingly, the magnetic ordering
abruptly disappears (inset of Fig.~\ref{fig:Ca1}~(a)).

In contrast, the BaFe$_2$As$_2$ Fermi surface shows a shrinkage of
the hole cylinders around (0,0,$k_z$) into hole pockets while the
Fermi surface sheets around ($\pi/a,\pi/a,k_z$) are changed
insignificantly (see Fig.~\ref{fig:BSFS2}). Therefore, as the
shrinkage occurs under pressure, the vanishing of Fermi surface
nesting at the lower $k_z$-plane is partially compensated by the
appearance of new nestings at a higher $k_z$-plane (e.g.
Fig.~\ref{fig:BSFS2} at $P/P_c''$=0.17 and $P/P_c''$=0.97), which
prohibits the sudden change of magnetization since the nesting
remains significant below the critical pressure (e.g.
Fig.~\ref{fig:BSFS2} at $k_z=2\pi/c$ and $P/P_c''$=0.97). Since
lowering the total energy by magnetic ordering always prevails over
abrupt lattice distortions, the sudden jump of the angles is
suppressed as shown in Fig.~\ref{fig:Ba}~(c). The slow change of the
Fermi surface topology as shown in Fig.~\ref{fig:BSFS2} is
consistent with the continuous reduction of magnetization as
observed in the inset of Fig.~\ref{fig:Ba} (a).

Fig.~\ref{fig:SUS} presents the calculated noninteracting
susceptibility $\chi_0 (q)$ at $q=(\pi,\pi,0)$ as a function of
pressure within the constant matrix element
approximation~\cite{Goldman2,Dong,Mazin2,Han2} for all the three 122
compounds. The relation between the change of the Fermi surface
nesting features  at $q=(\pi,\pi,0)$ and the phase transitions under
pressure is clearly quantified with this magnitude.
 $\chi_0 (q)$ at $q=(\pi,\pi,0)$ is
the response for the observed Fermi-surface-nesting-induced stripe-type
antiferromagnetic ordering and while it shows a continuous decrease at elevated
pressures in BaFe$_2$As$_2$, abrupt reductions at the phase transition
are detected in both CaFe$_2$As$_2$ and SrFe$_2$As$_2$, indicating
that the phase transition for these two systems are of first order where
a sudden
disappearance of the Fermi surface nesting occurs. Moreover, we
observe that the volume and lattice parameter discontinuities
(Fig.~\ref{fig:Sr}~(a) (b)) and the tetrahedron distortion
(Fig.~\ref{fig:Sr}~(c)) right above the phase transition for SrFe$_2$As$_2$
are of the
same nature as in CaFe$_2$As$_2$ but less pronounced. We conclude
that the phase transition is of weak first order in SrFe$_2$As$_2$.

Finally, we find that the ground state energy of the stripe-ordered
AF state is always far below that of the checkerboard-ordered one in
the orthorhombic phase while that of the high-pressure tetragonal
phase is non-magnetic, indicating that the phase transition under
pressure for the three compounds is not driven by the competition
between different magnetic ordering, {\it i.e.}, spin frustration.
Furthermore, the tiny crystal field splittings of the Fe 3d states
and strong itinerancy of all Fe 3d electrons does not support the
Kondo scenario~\cite{Vojta} for the phase transitions under
pressure.

In conclusion, we investigated the phase transitions in the iron
pnictide 122 compounds under pressure within the framework of {\it ab
  initio} molecular dynamics. The results on CaFe$_2$As$_2$ agree very
well with the experimental findings. We show that magnetic and
structural phase transitions also appear concomitantly in
SrFe$_2$As$_2$ and BaFe$_2$As$_2$. While they are weakly first order
in SrFe$_2$As$_2$, the phase transition is continuous in
BaFe$_2$As$_2$ which may be closely related to the different
behavior of entering superconductivity between SrFe$_2$As$_2$ and
BaFe$_2$As$_2$ observed experimentally~\cite{Alireza}. The remaining
hole pocket as well as nesting features in BaFe$_2$As$_2$ may be
related to the highest transition temperature~\cite{Alireza}
observed among 122 compounds due to the remaining spin fluctuations.
Since the topology of the Fermi surface in the high pressure phase
is quite different among these 122 compounds and
LaOFeAs~\cite{Opahle}, the pairing mechanism may differ among these
compounds although they belong to the same family of
superconductors.

We thank Claudius Gros for discussions and the Deutsche
Forschungsgemeinschaft for financial support through the TRR/SFB~49
and Emmy Noether programs and we acknowledge support by the
Frankfurt Center for Scientific Computing.

\end{document}